\documentclass[aps,twocolumn,prd,showpacs,nofootinbib]{revtex4}
\usepackage{amsmath}
\usepackage{graphicx}
\usepackage{dcolumn}
\usepackage{bm}
\usepackage{amssymb}
\usepackage{latexsym}

\def\be{\begin{equation}}
\def\ee{\end{equation}}
\def\ba{\begin{eqnarray}}
\def\ea{\end{eqnarray}}

\newcommand{\bd}{\ensuremath{\overline{\textnormal{D3}}}}
\newcommand{\dd}{\ensuremath{\overline{\textnormal{D}}}}

\begin{document}

\title{
D\dd ~Dark Energy  in String Warped Compactification
}

\author{Yun-Song Piao$^{a,b}$}
\affiliation{${}^a$Institute of High
Energy Physics, Chinese Academy of Sciences, P.O. Box 918-4,
Beijing 100039, P. R. China} \affiliation{${}^b$Interdisciplinary
Center of Theoretical Studies, Chinese Academy of Sciences, P.O.
Box 2735, Beijing 100080, China}

\begin{abstract}

We study the evolution of relic D3-branes after the D3/\bd-brane
inflation in string warped compactification. The motion of
D3-branes can be frozen under certain condition during the
radiation/matter domination.
These D3-branes can not be released until the D3/\bd-branes
potential energy becomes dominated at late time. Subsequently they
will move towards to \bd-branes, which play the role of uplifting
AdS minimum to dS minimum, near the apex of throats. The
annihilation of D3/\bd-branes leads to the disappearance of dS
vacua. This process may be regarded as a rapid decay channel of
present dS vacua. We discuss the parameter spaces required by this
process and calculate the decay time.

\end{abstract}

 \maketitle

The current set of cosmological data \cite{Bennett, Spergel,
Riess} implies that the universe might experience several separate
stages of acceleration, in each which the universe is in a near dS
state. The earlier accelerated stage driven by inflaton began
about 13 billion years ago, followed by a reheating process, in
which the universe is reheated to a suitable temperature. While
the recent accelerated expansion began approximately 5 billion
years ago, which is driven by dark energy constituting about 2/3
of the total energy density of the present universe.

Both stages of near dS expansion of universe have still challenged
string theory \cite{HKS, DKS, BFP}. It might become increasingly
clear that dS state could not be stable in any theory of quantum
gravity, instead it should be a metastable resonance with its
lifetime less than the recurrence time \cite{GKS}. Recently, there
have been some progresses on constructing the metastable dS vacua
in string theory. Based on the warped flux compactification
studied by GKP \cite{GKP} in type IIB string theory,
KKLT \cite{KKLT} stabilized the volume modulus in a supersymmetric
AdS minimum by taking into account nonperturbative effects, and
then they uplifted this AdS minimum to a metastable dS minimum by
adding some positive energy density contributions from \bd-branes
(or D7-branes with 2-form fluxes on its world-volume \cite{BKQ}).
The position of the dS minimum and the value of the cosmological
constant depend on the branes and on the quantized values of
fluxes in the bulk. In this frame, the early acceleration
(inflation) has been studied \cite{KKLMMT} in details.
D3/\bd-brane inflation \cite{BMN} can be expected to occur in a
warped throat \cite{Warped}. The distance between D3/\bd-brane is
regarded as inflaton and the small warped factor plays a
significant role for supporting inflation. But the further studies
in Ref. \cite{KKLMMT} have shown that when the details of
compactification are included, the D3-branes will obtain an
additional mass which is too big to allow for the occurrence of
inflation.

Generically, all dS compactifications in string theory could decay
and decompactify because the 10-dimension Poincare-invariant
string vacuum is supersymmetric and thus has vanishing energy
density. However, this is not the only decay mode. For example, in
any compactification in which RR fluxes contribute to the
potential energy, D-brane instantons can change the fluxes and
thus the value of the cosmological constant \cite{BP, FMSW}; while
in any compactification in which \bd-branes contribute to the
potential energy, as in KKLT, NS5-brane instantons can reduce the
NSNS flux and turn the \bd-branes into supersymmetric D3-brane
\cite{KPV, FLW}, and thus change the vacuum energy.

In this note, we study the evolution of relic D3-branes in the
bulk after the D3/\bd-brane inflation in KKLT-like
compactification. We are trying to give an interesting model in
which D3/\bd-branes annihilation induces to the decay of present
dS vacua. We assume that the inflaton shift symmetry \cite{HKP},
see also \cite{FT, HK}, or some degree of fine tuning is required
to set off the additional mass term from the compactification,
which will make the potential of D3/\bd-branes so flat that the
recent acceleration of observable universe can occur. The motion
of D3-branes can be frozen during the radiation/matter domination,
which may be realized by placing \bd-branes in a strongly warped
throat. They can not be released until the D3/\bd-branes potential
energy dominates the universe at late time.
We discuss the parameter spaces required by this process and
calculate the decay time.


{\it \bf Background-} We firstly review the background on which
our work is based. The GKP compactification \cite{GKP} of IIB
string theory on a threefold M with 7-branes and O3-planes can be
efficiently described as F theory \cite{V} compactification on a
CY fourfold X. 3-form fluxes and D3-branes added are subject to
the global tadpole constraint or the global conservation of RR
5-form ${\cal F}_5$ flux \be \label{tadpole} N_{\textnormal{D3}} -
N_{\bd} + \frac{1}{2\kappa_{10}^2 T_3} \int_{\textnormal{M}} {\cal
H}_3 \wedge  {\cal F}_3 - \frac{\chi ({\textnormal{X}})}{24}=0\ ,
\ee where $\kappa^2_{10}=(2\pi)^7(\alpha^\prime)^4 g_s^2/2$ is the
10-dimension Planck scale and $T_3 =1/(2\pi)^3(\alpha^\prime)^2
g_s$ is the 3-brane tension. The Euler number of the CY fourfold
$\chi({\textnormal{X}})$ gives the effective negative D3-branes
charge in IIB string theory of O3-planes and D7-branes wrapped on
4-cycles of M. This must be balanced by the charge from
4-dimension space-filling D3/\bd-branes, the wrapped NSNS and RR
3-form fluxes ${\cal H}_3$ and ${\cal F}_3$, which also source
${\cal F}_5$.

Following KKLT \cite{KKLT}, the non-perturbative correction ${\cal
A}e^{-a\rho}$ of ${\cal W}$ is introduced, which makes it (${\cal
W} ={\cal W}_0 + {\cal A}e^{-a\rho} $) fix the overall K\"{a}hler
modulus, in the meantime stabilize the volume modulus $\rho=\sigma
+i\alpha$ in a finite and moderately large value with a
supersymmetric AdS minimum $\Lambda_{\textnormal{AdS}} $. Then
they add some \bd-branes. These added \bd-branes can be balanced
in the tadpole condition (\ref{tadpole}) by turning on more
fluxes. The \bd-branes break the supersymmetry and bring an
additional term \be {\cal V} = \Sigma_i 2 p_i \beta^4_i T_3
\label{v0} .\ee to the effective potential of volume modulus
$\sigma$, In Planck unite, $T_3/m_p^4\sim 1/V_6^2\sim 1/\sigma^3$
\cite{KPV}, where $V_6$ is the warped volume of compactification
manifold, and $\beta$ is the redshift factor, which can be related
to the deformation parameter of the conifold tip and given by
$\beta^4 \sim \sigma \exp{(-{8\pi K\over 3 g_s M})}$. Thus Eq.
(\ref{v0}) gives a term $\sim 1/\sigma^2$, which leads to an
uplift to a metastable dS vacuum.

The background fluxes $M, K$ are given by \be M = {1\over 4\pi^2
\alpha^{\prime}} \int_{\textnormal{A}} {\cal F}_3 \ ,\ \ K =
-{1\over 4\pi^2 \alpha^{\prime}} \int_{\textnormal{B}} {\cal
H}_3\label{fluxquant}, \ee respectively, where the A cycle is the
$S^3$ which is finite at the tip, while the B cycle is the
6-dimensional dual of A. To minimize their energy, the \bd-branes
have to migrate to the apex of throats, so the energy density per
\bd-brane added depends on the fluxes. The $i$ in Eq. (\ref{v0})
labels the different throats. There could be large number of
3-cycles in a typical CY manifold, each of which can carry
nontrivial fluxes, so the existence of many throats could be
expected to be quite generic. In the background with multiple
throats the discrete density of vacua can increase dramatically.
By adjusting the fluxes in each individual throat, one may tune
${\cal V}$ with very high accuracy. Thus for sufficiently fine
tuning parameters, the additional term (\ref{v0}) in the potential
${\cal V}(\sigma)$ may lift the AdS global minimum to a required
local dS minimum.
The cosmological applications
of the compactification with multiple throats have been considered
in Ref. \cite{IT, C, BBC} for various aims.

{\it \bf D\dd ~Dark Energy-} We start with many wandering
D3/\bd-branes in the bulk. The \bd-branes can feel a net radial
force proportional to the 5-form flux ${\cal F}_5$, $F_r(r) = -2
T_3 {\cal F}_5 (r)$, and thus will be driven to the IR ends of
different throats. This force is a sum of gravitation and 5-form
contributions. For the D3-branes, both terms cancel. Thus though
during random transfer some of D3/\bd-branes can annihilate, the
case that many \bd-branes accumulate rapidly near the apex of
different throats may be still expected from a generic initial
distribution of D3/\bd-branes. These accumulated \bd-branes can
further attract D3-branes in the bulk, which under certain
conditions results in the acceleration of different stages of
observable universe.

The inflation of early universe may occur in a throat with a
weakly warped factor, inflation throat, in which the D3-branes in
the bulk will be driven and move towards to the \bd-branes near
the apex of inflation throat. The D3/\bd-branes after the
D3/\bd-brane inflation annihilate by the tachyon condensation and
their energy is released into the Standard Model degrees of
freedom of observable universe, see Ref. \cite{BBC} for a warped
reheating.
The universe after the reheating may have a KKLT-like vacuum with
its value equal to observed cosmological constant, and will
experience successive periods of radiation, matter and dark energy
domination following the evolution of standard cosmology.

The motion of D3-branes during the radiation/matter domination can
be different from that during the D3/\bd-brane inflation.
From the 4-dimension effective viewpoint, the rolling of a scalar
field can be frozen when the Hubble parameter during the
radiation/matter domination is larger than the effective mass of
this scalar field.
Thus similarly in this sense, the motion of D3-branes towards to
\bd-branes can become important only after their potential energy
starts to dominate the universe at late time.
During the D3/\bd-branes potential energy domination, the
D3-branes will be driven and annihilate \bd-branes near the apex
of throat with a strongly warped factor, dark energy throat. This
dark energy source is called as D\dd ~dark energy here. The dS
vacua after the annihilation of D3/\bd-branes will have lower
value or disappear, see Fig.1 for an illustration.

\begin{figure}[t]
\begin{center}
\includegraphics[width=8cm]{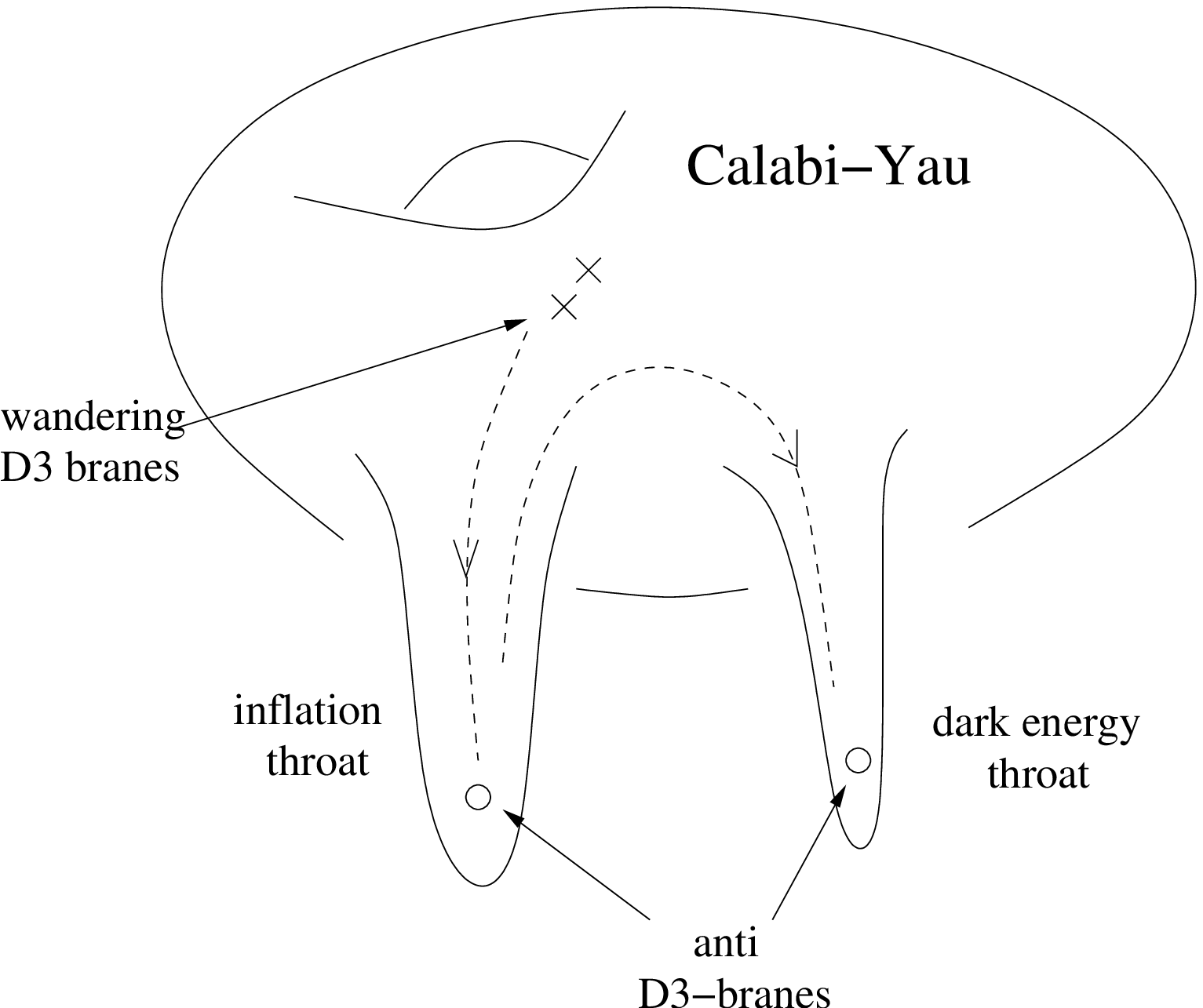}
\caption{ The illustration of D\dd ~dark energy in CY space.
\bd-branes are placed in the apex of throats, while some D3-branes
wander in the bulk. Initially D3-branes will be driven and move
towards to \bd-brane in inflation throat, after D3/\bd-branes
annihilation, the relic D3-branes will be attracted by \bd-branes
in dark energy throat, and rapidly move out of the inflation
throat and into the dark energy throat, subsequently the motion of
D3-branes is frozen due to the expansion of
radiation/matter-dominated universe. They can not be released
until the D\dd ~potential energy starts to dominate the universe.
}
\end{center}
\end{figure}

The potential between a D3-brane and $p$ \bd-branes in a warped
background takes the form \be {\cal V}(r)= 2p \beta^4 T_3 (1-
{1\over 2\pi^3} {\kappa^2_{10} \beta^4 T_3\over r^4}) , \label{v}
\ee where $r$ is the distance between D3/\bd-branes, $\beta$ is
the redshift factor $\sim \exp{(-{2\pi K\over 3 g_s M}) }$ and is
denoted by $r_0/R$ in Ref. \cite{KKLMMT}, and for conventions,
$(1/2\pi^3)\kappa^2_{10} T_3 =R^4/N = 4\pi g_s(\alpha^\prime)^2$
has been set here.

To make the kinetic term canonical, we implement the change
$\varphi = {\sqrt{T_3}} r$. The potential (\ref{v}) can be
rewritten as \be {\cal V}(\varphi) = 2p\beta^4 T_3 (1-{1\over
2\pi^2}{\beta^4 T_3\over \varphi^4}) ,\label{v2} \ee where
$\kappa^2_{10} T_3^2 =\pi $ has been used. For an uplift to a dS
state with its value equal to that of observed cosmological
constant, \be \Lambda(\varphi) ={\cal V}(\varphi)-
|\Lambda_{\textnormal{AdS}}|\simeq\Lambda_0 \label{lambda}\ee is
required, where $\Lambda_0$ is the present value of cosmological
constant of observable universe. To make the relic D3-branes not
have an opportunity to annihilate \bd-branes up to today, the
condition \be {{\Lambda}^{\prime\prime}\over h_0^2} \sim m_p^2
{{\Lambda }^{\prime\prime}(\varphi)\over \Lambda_0} \sim {20
p\over \pi^2}{\beta^8 m_p^2 T_3^2\over \Lambda_{0} \varphi^6} \leq
1 \label{eta}\ee must be satisfied, where $h_0^2\sim
\Lambda_0/m_p^2$ is the present Hubble parameter with $m_p=T_3
V_6^{1\over 2}/\sqrt{\pi}$ being 4-dimension Planck scale. Eq.
(\ref{eta}) gives \be {20 p\over \pi^3 }{\beta^8 T_3 \over
\Lambda_0} \leq {r^6\over V_6} < 1 ,\label{r6} \ee where the
reason of the latter inequality is that two branes can not be
separated by a distance larger than the volume of compactification
manifold.
For $|\Lambda_{\textnormal{AdS}}|\gg \Lambda_0$, $ {\cal
V}(\varphi)\simeq 2p\beta^4 T_3 \simeq
|\Lambda_{\textnormal{AdS}}|$ (for
$|\Lambda_{\textnormal{AdS}}|\simeq \Lambda_0 $, $ 2p\beta^4 T_3
\simeq \Lambda_{0}$ is also suitable for replacing) can be
obtained, thus \be \beta^4 \simeq
{|\Lambda_{\textnormal{AdS}}|\over 2p T_3} . \label{beta8}\ee
Instituting (\ref{beta8}) into (\ref{r6}), we have \be p> {5\over
\pi^3}{|\Lambda_{\textnormal{AdS}}|^2\over T_3 \Lambda_0} .
\label{p}\ee Thus for $5|\Lambda_{\textnormal{AdS}}|^2\leq \pi^3
T_3 \Lambda_0$, one or several \bd-branes will be enough for the
uplifting to a dS minimum with its value equal to that of observed
cosmological constant, while for
$5|\Lambda_{\textnormal{AdS}}|^2\gg \pi^3 T_3 \Lambda_0$, a large
number of \bd-branes will be required, see the upper right panel
of Fig.2 for a numerical analysis. Notice that the potential
(\ref{v}) between a D3-brane and $p$ \bd-branes is valid only for
small $p$, while for enough large $p$, the correction to
background will become important.


\begin{figure}[t]
\begin{center}
\includegraphics[width=8cm]{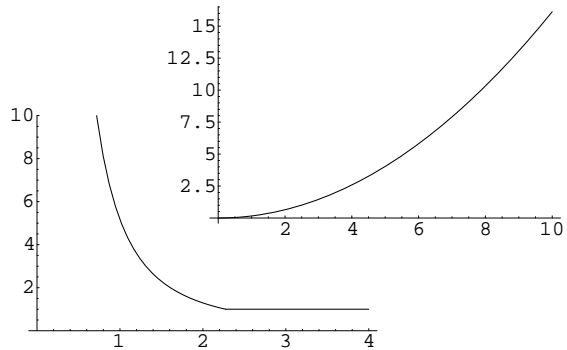}
\caption{ In the upper right panel, the x-axe is
${|\Lambda_{\textnormal{AdS}}|\over \sqrt{T_3 \Lambda_0}} $, and
the y-axe is $p$. The solid is the lower limit of $p$, the
integral points above which are the value of $p$ making the
inequality (\ref{p}) valid. In the lower left panel, the x-axe is
${|\Lambda_{\textnormal{AdS}}|\over \sqrt{T_3 \Lambda_0}} $, and
the y-axe is ${T\over T_0}$. The solid is the upper limit of
${T\over T_0} $, }
\end{center}
\end{figure}

We then calculate the decay time $T$ of D\dd ~dark energy in small
$p$ limit.
In principle, all wandering D3-branes during the D3/\bd-brane
inflation will be attract into the inflation throat, and after the
D3/\bd-brane inflation the relic D3-branes will move out of the
inflation throat and into the dark energy throat. To make this
process feasible, the warping of inflation throat should be weaker
than that of dark energy throat.
Thus the time of exiting from the inflation throat, compared with
the time of staying in the dark energy throat, can be neglected.
Two points may be further considered for this calculation. One is
that during the radiation/matter domination, the motion of
D3-branes is nearly frozen due to the expansion of universe, which
provides an initial value for time integral. The other is that
when the slow rolling condition (\ref{eta}) is broke down,
D3-branes will be expected to annihilate \bd-branes rapidly. Thus
the majority of the moving time should be from the slow rolling
process of D3-brane, in which \be 3h{\dot \varphi} + {\Lambda
}^{\prime}(\varphi)\simeq 0 \ee can be satisfied.
Thus \be T\simeq \int dt \simeq -\int {3h\over {\Lambda
}^{\prime}(\varphi) } d\varphi . \label{t} \ee Instituting
$h^2\simeq h_0^2 = \Lambda_0 /3m_p^2 $ and Eq. (\ref{lambda}) into
Eq. (\ref{t}), we have \ba T &\simeq &-\int {\pi^3\over 4\sqrt{3}
p }{\sqrt{\Lambda_0} \over m_p \beta^8 T_3^2} \varphi^5 d\varphi
\nonumber\\ &=& {\pi^3\over 24\sqrt{3} p } {\sqrt{\Lambda_0}
T_3\over m_p \beta^8 } r^6 |^r_{r_0} \nonumber\\ &<& {\pi^3
\sqrt{\pi}\over 24\sqrt{3} p } {\sqrt{\Lambda_0} V_6^{1\over 2}
\over \beta^8 }\sim {\cal O}(1) {\sqrt{\Lambda_0} V_6^{1\over 2}
\over p \beta^8 } , \label{t2} \ea where $T_3 V_6^{1\over 2}=
\sqrt{\pi} m_p$ and $r^6 < V_6$ has been used. Further $\beta$ can
be cancelled by using Eq. (\ref{beta8}), we obtain \be T < {\cal
O}(1) {p T_3^2 V_6^{1\over 2} \Lambda_0^{1/2}\over
|\Lambda_{\textnormal{AdS}}|^{2}} . \label{t3}\ee For a compare,
defining $T_0\simeq 1/h_0 = \sqrt{3}m_p/ \sqrt{\Lambda_0} $ as the
present age of observable universe, we have \be {T\over T_0}<
{\pi^3\over 6} {p \Lambda_0 T_3 \over
|\Lambda_{\textnormal{AdS}}|^{2}} . \label{ttc}\ee Thus for
$6|\Lambda_{\textnormal{AdS}}|^2 > \pi^3 T_3 \Lambda_0$, D\dd
~dark energy will decay in a cosmological age, while for
$6|\Lambda_{\textnormal{AdS}}|^2\leq \pi^3 T_3 \Lambda_0$, the
decay time of D\dd ~dark energy will be many times longer than the
age of universe. The lower left panel of Fig.2, after we combine
(\ref{p}) and (\ref{ttc}), is plotted, in which only lower limit
of $p$ is considered, {\it i.e.} for $x\equiv
{|\Lambda_{\textnormal{AdS}}|\over \sqrt{T_3 \Lambda_0}} \geq
\pi^3/6$, $p=1$ is taken and for $x\leq \pi^3/6$, $p$ is
approximately $(5/\pi^3)x^2$. For larger value of $p$, the solid
line of Fig.2 will rise, and thus the decay time will longer.

{\it \bf Discussion- }In KKLT-like compactification, many
\bd-branes are placed in different throats and required to
implement the uplifts from AdS minimum to dS minimum. For a
generic initial distribution, some D3-branes after the
D3/\bd-brane inflation may be left. These relic D3-branes will be
driven and annihilate \bd-branes near the apex of dark energy
throats, which may induce a rapid decay of present dS vacua.

We study this process in this note. The main point is that the
motion of D3-branes can be frozen during the radiation/matter
domination, which may be realized by placing \bd-branes in a
strongly warped throat, in which
${\Lambda}^{\prime\prime}(\varphi)< h_0^2$ can be ensured. But the
parameter spaces for the number $p$ of \bd-branes and the energy
scale $|\Lambda_{\textnormal{AdS}}|$ of AdS minimum are not
arbitrary. We find that only for
$5|\Lambda_{\textnormal{AdS}}|^2\leq \pi^3 T_3 \Lambda_0$, one or
several \bd-branes are suitable for the uplift to an observed
value of cosmological constant, while for a larger
$|\Lambda_{\textnormal{AdS}} |$, more \bd-branes will be required,
otherwise D3/\bd-branes will annihilate so early that we can not
observe the D\dd ~dark energy. In a model with multiple dark
energy throats, whether can one find other uplifting modes of
using \bd-branes to relax above conditions? For example, one may
firstly uplift AdS minimum to another AdS minimum with lower
absolute value by placing some \bd-branes in a throat with a
weakly warped factor, then uplift this new AdS minimum to dS
minimum by same step but in a throat with a strongly warped
factor. However, it seems that this uplifting mode will make the
thing worse. Generally, the weakly warped factor will lead to the
strong attraction, thus D3-branes will be firstly driven into
these throats with the weakly warped factors and annihilate
\bd-branes in these throats.

Generally, D\dd ~dark energy will inevitably lead to a sharp
change of the value of dark energy in the future, either reducing
its value or leading to a catastrophic decay to AdS state (The
negative cosmological constant will eventually induce a Big Crunch
\cite{FFKL}, see also Ref. \cite{Piao} for a cyclic universe
experiencing different AdS minima.).
The decay time of D\dd ~dark energy is faster than that of
decompactification to 10-dimension Minkowski vacua \cite{KKLT} and
NS5-brane mediated decay \cite{KPV, FLW}. For a larger
$|\Lambda_{\textnormal{AdS}}|$, its decay time is almost the same
order as the present age of universe.
Further, it may be interesting to compare D\dd ~dark energy model
with current observational constraints on cosmic doomsday
\cite{WKLS}. Finally, of course one can also specially design some
models which evade the decay channel proposed here, for example,
by simply not having relic D3-branes after the D3/\bd-brane
inflation, as in KKLT \cite{KKLT},
Thus in this case one of the other decay channels to a
10-dimension Minkowski vacuum or a different flux vacuum will have
to occur.

{\bf Acknowledgments} The author would like to thank Andrew Frey,
Shamit Kachru for helpful comments, and Henry Tye for criticisms.
This work is supported in part by K.C. Wang Postdoc Foundation,
also in part by NNSFC under Grant No: 10405029. 90403032 and by
National Basic Research Program of China under Grant No:
2003CB716300.

\end{document}